# Long-Term Evolution of Email Networks: Statistical Regularities, Predictability and Stability of Social Behaviors

Antonia Godoy-Lorite[1], Roger Guimerà[1,2]*, Marta Sales-Pardo[1]

**1** Departament d'Enginyeria Química, Universitat Rovira i Virgili, 43006 Tarragona, Catalonia, Spain,
**2** Institució Catalana de Recerca i Estudis Avançats (ICREA), 08010 Barcelona, Catalonia, Spain

\* roger.guimera@urv.cat

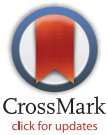

## Abstract

In social networks, individuals constantly drop ties and replace them by new ones in a highly unpredictable fashion. This highly dynamical nature of social ties has important implications for processes such as the spread of information or of epidemics. Several studies have demonstrated the influence of a number of factors on the intricate microscopic process of tie replacement, but the macroscopic long-term effects of such changes remain largely unexplored. Here we investigate whether, despite the inherent randomness at the microscopic level, there are macroscopic statistical regularities in the long-term evolution of social networks. In particular, we analyze the email network of a large organization with over 1,000 individuals throughout four consecutive years. We find that, although the evolution of individual ties is highly unpredictable, the macro-evolution of social communication networks follows well-defined statistical patterns, characterized by exponentially decaying log-variations of the weight of social ties and of individuals' social strength. At the same time, we find that individuals have social signatures and communication strategies that are remarkably stable over the scale of several years.

**Data Availability Statement:** Data are available from Figshare (http://dx.doi.org/10.6084/m9.figshare.1577586).

**Funding:** This work was supported by a James S. McDonnell Foundation Research Award, Spanish Ministerio de Economía y Competitividad (MINECO) Grants FIS2010- 18639 and FIS2013-47532-C3, European Union Grant PIRG-GA-2010-277166 (to RG), European Union Grant PIRG-GA-2010-268342 (to MSP), and European Union FET Grant 317532 (MULTIPLEX). The funders had no role in study design, data collection and analysis, decision to publish, or preparation of the manuscript.

## Introduction

Individuals thrive in a social environment through the construction of social networks. Ties in these networks satisfy individual needs and are necessary for well-being, but the effort, time and cognitive investment that each tie requires limit the ability of individuals to maintain them [1–3]. As a result of this limit, social networks are intrinsically dynamical, with individuals constantly dropping ties and replacing them by new ones [1, 2, 4].

Several factors are known to play an important role in the intricate microscopic process of tie replacement—for example, mechanisms such as homophily [5] and triadic closure [6] have been found to generally drive tie creation [4]. However, these processes are remarkably noisy [4] and are modulated by the distinct social behaviors of each individual [1–3], so that in the short term individual ties appear and decay in a highly unpredictable fashion.

Here we investigate whether, despite the intricacies and randomness of the tie formation and decay processes at the microscopic level, there are macroscopic statistical regularities in the long-term evolution of social communication networks. Statistical regularities have indeed been reported in the activity patterns of single individuals, and are likely driven by daily and weekly periodicities (e.g. in communication [7–10] and mobility [11, 12]); statistical regularities have also been reported in the long-term evolution of human organizations [13–17] and human infrastructures such as the air transportation system [18]. However, due to the difficulty of tracking social interactions of a large pool of individuals for a long time, we still lack a clear picture of what statistical regularities emerge in the long-term evolution of social networks. In particular, beyond relatively short periods of time of 12 to 18 months [2–4, 19], we do not know up to what extent social networks remain stable, or whether individuals change their social behavior with time.

Besides the academic interest of these questions, they are also of practical relevance because the structure of social networks plays an important role in processes such as the spread of information or epidemics [20–22]. The static analysis of communication networks has shed light on some important aspects (e.g. the role of weak ties in keeping the stability of social networks [23]). However, it is increasingly clear that ignoring network dynamics can lead to very poor models of collective social behavior, and that even fluctuations at a microscopic level often have a large impact on social processes [24].

To elucidate these questions, here we analyze the evolution of an email network [25] of hundreds of individuals within an organization over a period of four consecutive years. We find that, although the evolution of individual ties is highly unpredictable even in the long term, the macro-evolution of social communication networks follows well-defined statistical patterns, characterized by exponentially decaying log-variations of the weight of social ties and of individuals' social strength. At the same time, we find that individuals have long-lasting social signatures and communication strategies.

## Data

We analyze the email network of a large organization with over 1,000 individuals for four consecutive years (2007-2010). For this period, we have information of the sender, the receiver and the time stamp of all the emails sent within the organization using the corporate email address. To preserve users' privacy, individuals are completely anonymized and we do not have access to email content (see Methods). The email networks for each year comprise $n_{2007} = 1,081$, $n_{2008} = 1,240$, $n_{2009} = 1,386$, and $n_{2010} = 1,522$ individuals. The total number of emails recorded each year is $l_{2007} = 211,039$, $l_{2008} = 303,619$, $l_{2009} = 368,692$, and $l_{2010} = 444,493$.

Since the number of emails sent from $i$ to $j$ during a year is typically similar to the number of emails sent from $j$ to $i$ (see S1 File), we consider the undirected weighted network in which the weight $\omega_{ij}$ of the connection between users $(i, j)$ represents the total number of emails exchanged by this pair of users during one year. Because we are interested in non-spurious social relationships, in our analysis we only consider connections with weight $\omega_{ij} \geq 12$, that is we only consider connections between pairs of users that exchange at least an email per month on average. Such filters are known to generate networks whose connections resemble more closely self-reported social ties [26].

## Results

### The long-term evolution of email communication follows well-defined statistical patterns

We characterize the long-term evolution of email communication networks in terms of two properties: the weight $\omega_{ij}(t)$ of connections for year $t$ (Fig 1); and the user strength $s_i(t) = \sum_j \omega_{ij}(t)$ (Fig 2) [27], that is, the total number of emails exchanged by each user $i$ during year $t$.

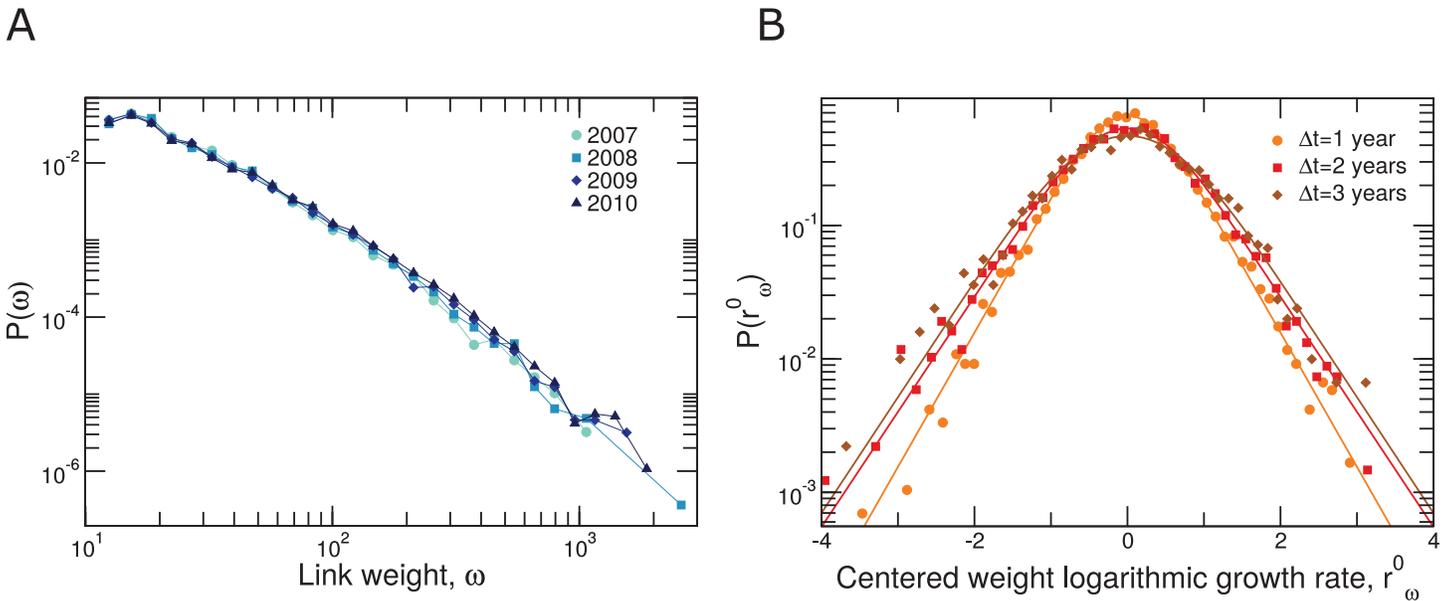

**Fig 1. Time evolution of connections' weights.** The weight $\omega_{ij}$ of a connection between users $(i, j)$ corresponds to the number of emails exchanged by $i$ and $j$ during a whole year. We only consider connections with $\omega \geq 12$ (see text) **(A)** Distributions of weights for each one of the years in our dataset (2007–2010). Note that the distribution is stable in time. **(B)** Distribution of the centered weight logarithmic growth rates $r_\omega^0 = log(\omega(t + \Delta t)) - log(\omega(t)) - \mu(t, \Delta t)$ for $\Delta t = 1, 2, 3$ (dots, squares and diamonds, respectively). Lines show fits to the convolution of a Laplace distribution and a Gaussian distributed noise (see Eq (5)) (parameters $\Delta t = 1$: $\sigma_{exp} = 0.43$, and $\sigma_G = 0.35$, $\Delta t = 2$: $\sigma_{exp} = 0.50$, and $\sigma_G = 0.47$ and $\Delta t = 3$: $\sigma_{exp} = 0.50$, and $\sigma_G = 0.60$). Note that as $\Delta t$ increases the peaks are rounder and the distributions are slightly wider (see Fig D in S2 File). See Fig B in S2 File for values of the distribution modes $\mu(t, \Delta t)$.

doi:10.1371/journal.pone.0146113.g001

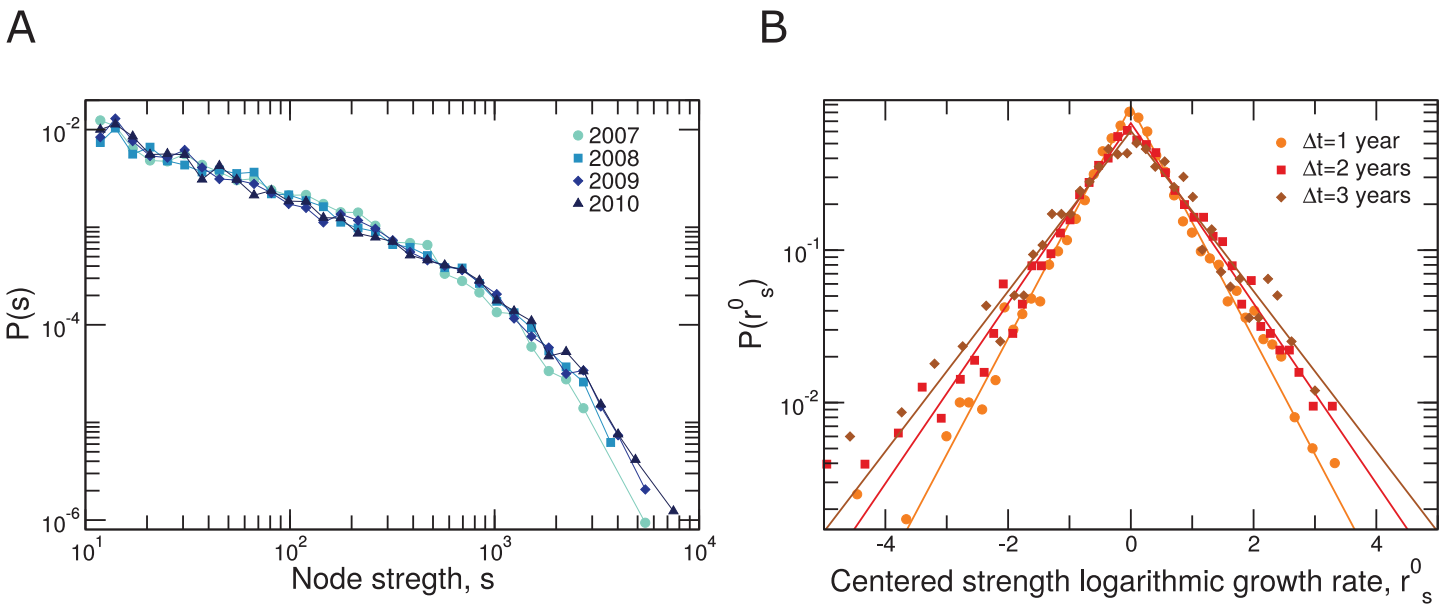

**Fig 2. Time evolution of nodes' strengths.** The strength $s_i$ of node $i$ is the number of emails that user $i$ exchanged with other users during one year. **(A)** Distributions of strengths for each one of the years in our dataset (2007–2010). Note that the distribution is stable in time. **(B)** Distribution of centered strength logarithmic growth rates $r_s^0 = log(s(t + \Delta t)) - log(s(t)) - \mu(t, \Delta t)$ for $\Delta t = 1, 2, 3$ years (dots, squares and diamonds, respectively). Lines show fits to a Laplace distribution (parameters $\Delta t = 1$: $\sigma_{exp} = 0.57$, $\Delta t = 2$: $\sigma_{exp} = 0.74$ and $\Delta t = 3$: $\sigma_{exp} = 0.83$). Note that as $\Delta t$ increases the distributions are wider (see Fig D in S2 File). For the specific values of the distribution modes $\mu(t, \Delta t)$ see Fig B in S2 File.

doi:10.1371/journal.pone.0146113.g002

The distributions of connection weights and user strengths have two remarkable features (Figs 1A and 2A). First, these distributions are fat-tailed, with values spanning over three orders of magnitude. Second, these distributions are stable for the four years we study (despite a small but significant shift towards higher number of emails).

Besides the overall stability of the distributions, we observe a large variation in connection weights and user strengths from year to year. To characterize this variation, we define the logarithmic growth rates [13–18]

$$r_\omega(t, \Delta t) = \log\left(\frac{\omega(t + \Delta t)}{\omega(t)}\right) \qquad (1)$$

$$r_s(t, \Delta t) = \log\left(\frac{s(t + \Delta t)}{s(t)}\right), \qquad (2)$$

and study their distributions (Figs 1B and 2B). These distributions are tent-shaped and have exponentially decaying tails. For fixed $\Delta t$ the mode $\mu(t, \Delta t)$ of the distribution changes slightly with the starting year $t = 2007, 2008, 2009$, which is significant for $t = 2007$ but not significant for $t = 2008$ and $t = 2009$ (see Figs B and C in S2 File). Remarkably, if we consider the distributions of logarithmic growth rates centered at zero $r^0 = r - \mu(t, \Delta t)$ then these distributions are stationary (see S2 File). Moreover, the same functional form that describes growth rates from one year to the next, $\Delta t = 1$ year, also describes growth rates at $\Delta t = 2$ years and $\Delta t = 3$ years. For user strengths, a Laplace distribution

$$P_L(r^0) = \frac{\exp\left(-|r^0|/\sigma_{\exp}\right)}{2\sigma_{\exp}} \qquad (3)$$

provides the best overall fit to the data (as determined by the Bayesian information criterion [28]; see S2 File). For connection weights a pure Laplace distribution does not provide a good fit to the data because of the rounding of the distribution around its mode. In this case, we obtain the best fit if we assume that the observed centered rate $r^0$ is a combination $r^0 = \tilde{r^0} + \epsilon$, where $\tilde{r^0}$ is Laplace distributed according to Eq (3) and $\epsilon$ is a normally distributed "noise", so that $P(r^0)$ is the convolution of a Laplace and a normal distribution (see Eq (5) and S2 File in Supplementary Material). Note that as $\Delta t$ increases, the width of the Laplace distribution and the intensity of the Gaussian noise in $P(r_\omega^0)$ increases.

Tent-shaped distributions with exponentially decaying tails are common in the growth of human organizations [13–17], and have also been reported in the growth of complex weighted networks [18]. The exponential tails of these distributions imply that fluctuations in connection weights and user strengths are considerably larger than one would expect from a process with Gaussian-like fluctuations.

## Logarithmic growth rates are largely unpredictable despite significant correlations

The fact that long-term growth rates follow well-defined distributions raises the question of whether it is possible to quantitatively predict the evolution of the network. To investigate this, we start by studying whether there are long-term trends in the logarithmic growth rates. Note that in the previous section we have found that logarithmic growth rate distributions are not stationary as distributions change slightly their modes (see Fig B in S2 File). However, we note that the displacement of the distribution is small compared to prediction errors. Additionally, since the aim is to predict future growth rates, the displacement of future distributions is in

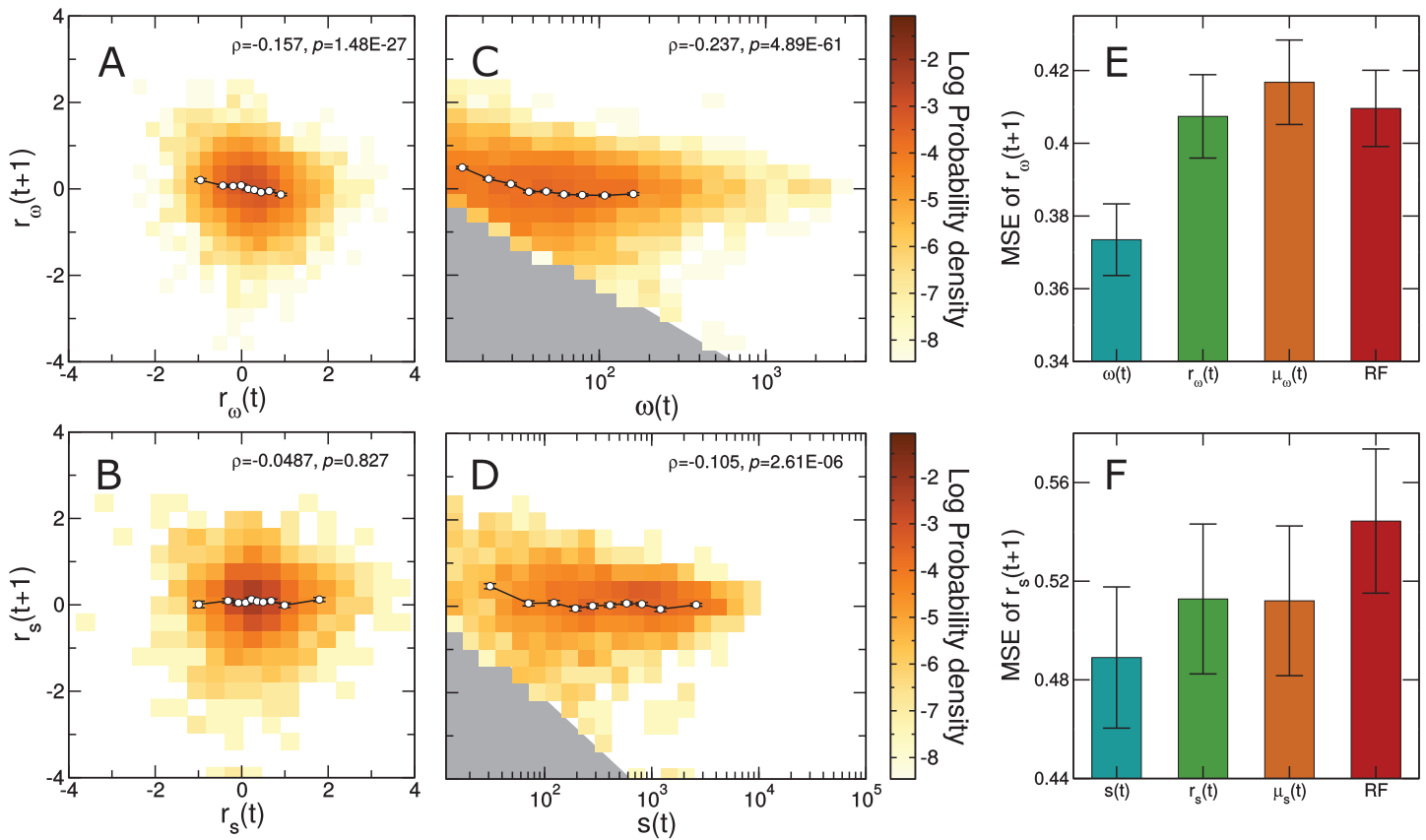

**Fig 3. Predictability of logarithmic growth rates for connection weight $r_\omega(t+1)$ (A, C, E) and user strength $r_s(t+1)$ (B, D, F).** (A) Joint probability density of $r_\omega(t+1)$, the logarithmic growth rate of weights at time $t+1$, and $r_\omega(t)$, the logarithmic growth rate of weights at time $t$. (B) Joint probability density of $r_s(t+1)$, the logarithmic growth rate of strengths at time $t+1$, and $r_s(t)$, the logarithmic growth rate of strengths at time $t$. (C) Joint probability density of $r_\omega(t+1)$, the logarithmic growth rate of weights at time $t+1$, and $\omega(t)$, the weight at time $t$. The area shaded in grey area is no allowed since $r_\omega(t+1) \geq -\log \omega(t)$. (D) Joint probability density of $r_s(t+1)$, the logarithmic growth rate of strengths at time $t+1$, and $s(t)$, the strength at time $t$. The area shaded in grey is forbidden since $r_s(t+1) \geq -\log s(t)$. In plots (A–D), circles and error bars show the mean and one standard error of the mean for values binned along the X axis. It is visually apparent that $\omega(t)$ and $s(t)$ are more informative about $r_\omega(t+1)$ and $r_s(t+1)$, respectively, than $r_\omega(t)$ and $r_\omega(t)$ (as confirmed by Spearman's $\rho$ and p-values, displayed inside each graph). (E, F) Root mean squared error (MSE) of the predictions of the logarithmic growth rates at time $t+1$ obtained from leave-one-out experiments. As predictors, we use: (E) $\omega(t)$, $r_\omega(t)$, and $\mu_\omega(t)$ (see Eq (5)); (F) $s(t)$, $r_s(t)$, and $\mu_s(t)$ (see Eq (3)). Additionally, in both cases we try to predict the logarithmic growth rate using a Random Forest regressor [29]. Note that a simple approach (i.e. considering the weight/strength at time $t$) performs significantly better than a well-performing machine learning algorithm such as the Random Forest. In any case, and despite being the most predictive, weight/strength at time $t$ only provide moderate improvements over predictions made using the mean value $\mu_\omega$ for all connections and $\mu_s$ for all users.

doi:10.1371/journal.pone.0146113.g003

practice unknown. Therefore we use uncentered logarithmic growth rates $r_\omega$ and $r_s$ in the prediction analysis. In particular, we analyze whether there are significant correlations between the logarithmic growth rate in one year and the logarithmic growth rate the following year (Fig 3A–3B). We find that the correlation is not significant for strength logarithmic growth rates, and significant but negative for weight logarithmic grow rates (Spearman's $\rho = -0.16$, $p = 1.5 \cdot 10^{-27}$).

In fact, we find that the network properties at time $t$ that are most correlated with the logarithmic growth rates $r_\omega(t+1)$ and $r_s(t+1)$ are the connection weight (Spearman's $\rho = -0.24$, $p = 4.9 \cdot 10^{-61}$) and the user strength (Spearman's $\rho = -0.11$, $p = 2.6 \cdot 10^{-6}$), respectively (Fig 3C–3D; see S3 File for other network properties). These correlations are negative, which indicates that small values of connection weight and user strength grow faster than large values, also that negative values of weights and strengths are not allowed. In any case, despite the significance of these correlations, the high variability of $r_\omega(t + 1)$ and $r_s(t + 1)$ for fixed values of $\omega(t)$ and $s(t)$, respectively, raises the question of whether the correlations can be used reliably to predict the evolution of the network.

To quantify the predictive power of these variables, we carry out leave-one-out experiments to predict logarithmic growth rates $r_\omega(t + 1)$ and $r_s(t + 1)$ from network properties at time $t$ (Fig 3E–3F). We investigate different approaches: (i) assuming the same growth equal to the mean growth $\mu_\omega(t)$ and $\mu_s(t)$ for all predictions of $r_\omega(t + 1)$ and $r_s(t + 1)$, respectively (note that as it is shown in Fig D in S2 File, mean growths are very close to zero); (ii) using individual network observables as predictors, in particular, $\omega(t)$ and $r_\omega(t)$ for $r_\omega(t + 1)$, and $s(t)$ and $r_s(t)$ for $r_s(t + 1)$; and (iii) using a well-performing machine learning approach such as a Random Forest regression [29] with an array of network observables (see S3 File for more details). We find that using the Random Forest does not yield significantly better predictions than using the average expected growth for all predictions. Using the most correlated variables $\omega(t)$ and $s(t)$ for $r_\omega(t + 1)$ and $r_s(t + 1)$ respectively, only shows a modest improvement (Fig 3E–3F). Our results therefore suggest that the existence of correlations is not enough to build a satisfactory predictive model for the logarithmic growth rates (and that black box methods like Random Forests may, in fact, be even less appropriate).

## Social signatures are stable in the long term

Next, we seek to better understand the evolution of the communication behavior of individual users. Recent results suggest that the way individuals divide their communication effort among their contacts (their so-called "social signature") is stable over the period of a few months [3]. This is consistent with the hypothesis that humans have a limited capacity to simultaneously maintain a large number of social interactions [1, 30].

Here, we investigate whether social signatures are stable over the period of several years. In particular, we analyze how individuals distribute their communication activity (their emails) among their contacts. To quantify how evenly distributed emails are among those contacts, we use the standardized Shannon entropy $S_i$

$$S_i = \frac{-\sum_{j=1}^{k_i} \frac{\omega_{ij}}{s_i} \log \frac{\omega_{ij}}{s_i}}{\log k_i} , \qquad (4)$$

where $k_i$ is the number of contacts of user $i$. Note that $S_i = 1$ when user $i$ exchanges the same number of emails with all her contacts and $S_i \approx 0$ when she exchanges almost all of her emails with a single contact (Fig 4A). We use the standardized Shannon entropy because it shows a smaller dependence on the number of contacts than other measures of social signature such as the Gini coefficient (Fig E in S4 File).

We find that the distribution of standardized entropies is heavily shifted towards high values of $S_i$ (Fig 4B), which implies that most individuals tend to distribute their communication evenly among all their contacts. We also find that the overall distribution of social signatures is stable in time (see also Supplementary Material).

To study the stability of each individual's social signature, we measure the difference $\Delta S_i(\Delta t) = S_i(t + \Delta t) - S_i(t)$ for $\Delta t = 1, 2, 3$ years (Fig 4B). We find that the distribution of $\Delta S_i(\Delta t)$ is symmetric and heavily peaked around zero and stable for any fixed value of $\Delta t$ (Fig A in S4 File). Therefore since most of the users do not change their social signature during the three year period of our analysis, our results suggest that individual's social signatures are stable in the long term.

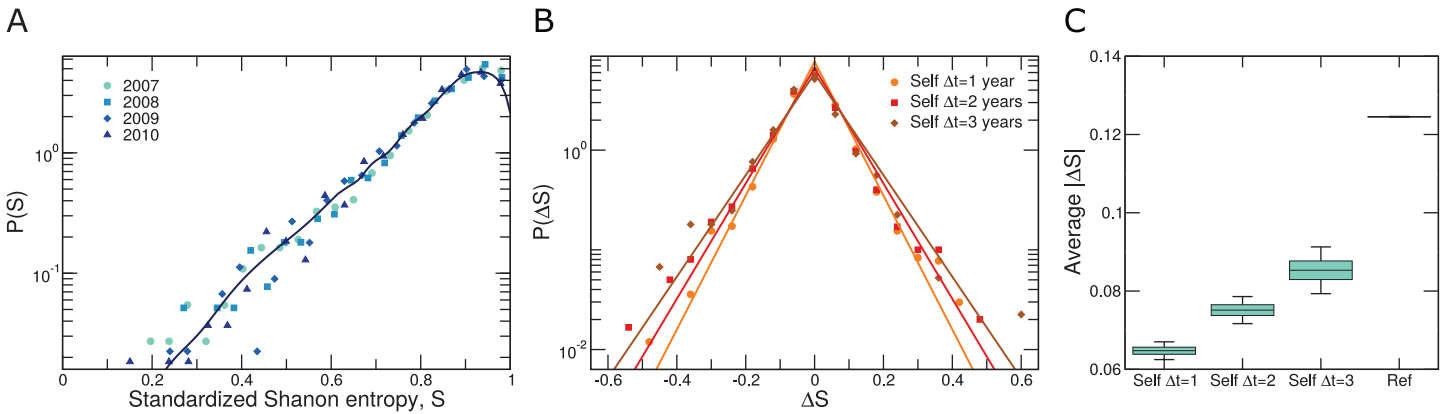

**Fig 4. Stability of social signatures.** (A) Distribution of the standardized Shannon entropy $S_i$ (see text) for users in the period 2007–2010. Entropy quantifies the extent to which and individual's communication efforts are distributed among her contacts, so that $S_i = 1$ when user $i$ exchanges the same number of emails with all her contacts and $S_i \approx 0$ when she exchanges almost all of her emails with a single contact. Distributions for all years collapse onto a single curve. The line shows a kernel density estimation of the four yearly datasets pooled together. (B) Distributions of the change of individual standardized Shannon entropy $\Delta S_i(\Delta t) = S_i(t + \Delta t) - S_i(t)$, $\forall i$ for $\Delta t = 1, 2, 3$ years (dots, squares and diamonds, respectively). The lines show the Laplace best fits based on BIC for the three distributions ($\Delta t = 1 \ \sigma = 0.065$; $\Delta t = 2 \ \sigma = 0.075$; and $\Delta t = 3\sigma = 0.085$). (C) Comparison between the absolute difference in individual social signatures $|\Delta S_i(\Delta t)|_{self} = |S_i(t + \Delta t) - S_i(t)|$ and the typical absolute difference of entropies between individuals $|\Delta S_{ij}|_{ref} = |S_i(t) - S_j(t)|$. The boxplot shows unambiguously that users have stable social signatures.

doi:10.1371/journal.pone.0146113.g004

To quantify this more precisely, we compare the absolute change of a user's standardized entropy $|\Delta S_i(\Delta t)|_{self} = |S_i(t + \Delta t) - S_i(t)|$ to the typical absolute difference of entropies between individuals $|\Delta S_{ij}|_{ref} = |S_i(t) - S_j(t)|$, $\forall j \neq i$ (Fig 4C). We observe that the variation of the social signature of a user in time is typically much smaller (even when $\Delta t = 3$ years) than the variation between individuals, confirming that the social signature is a trait of users that persists even during periods of several years (see Fig C in S4 File for the analysis of each individual year). In fact, by extrapolating the values of $|\Delta S_i(\Delta t)|_{self}$, we estimate that individual social signatures may be persistent for roughly eight years.

## Communication strategy is stable in the long term

A related question to the stability of the social signature is that of whether users tend to keep the same contacts over time or not. Recent studies have shown that, in the short term, individuals differ in their communication strategies [1]—some individuals tend to change their contacts frequently ("explorers"), whereas others tend to maintain contacts ("keepers"). We investigate whether these differences exist at the scale of years and if individual communication strategies are stable in the long term.

To that end, we consider the fraction $f_i(t)$ of all emails exchanged by user $i$ in year $t$ (out of the total $s_i(t)$) with preexisting contacts, that is users with whom user $i$ had also exchanged emails during the previous year, $t - 1$. Therefore, $f_i(t) = 1$ means user $i$ exchanged all her emails in year $t$ with preexisting contacts, whereas $f_i(t) = 0$ means that user $i$ only exchanged emails with new contacts.

The distribution of $f_i$ (Fig 5A) shows that most individuals are social keepers (see also Fig G in S4 File for the turnover of the network contacts). Indeed, the mode of the distribution is around $f_i(t) = 0.9$, and 58% of the users exchange more than 75% of their emails with preexisting contacts. Still, a non-negligible 17% of the individuals exchange more than half of their emails in one year with new contacts. Our findings thus confirm that, even at the scale of years, there is a variety of communication strategies [1].

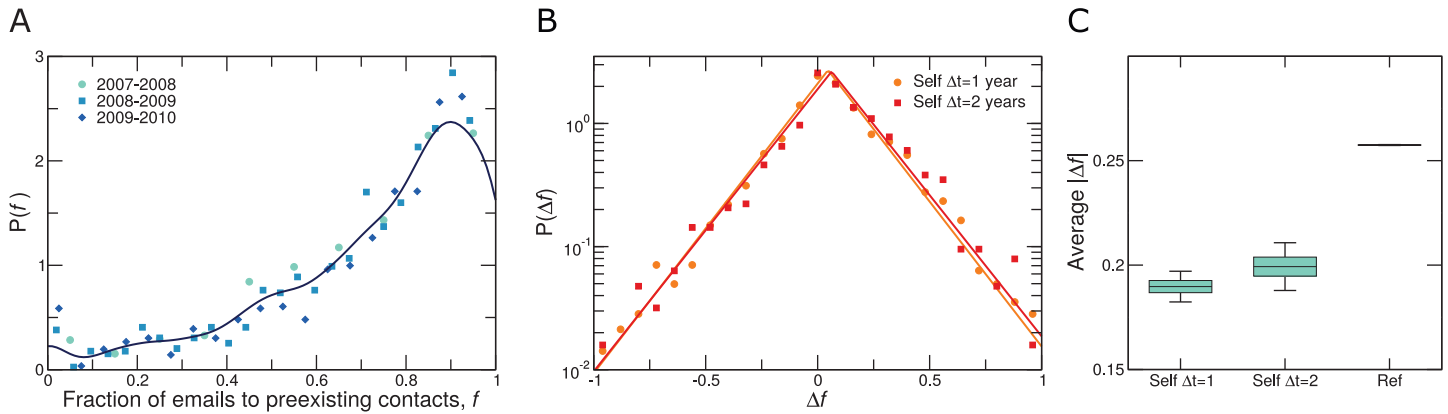

**Fig 5. Stability of individual communication strategies. (A)** Distribution of the fraction of emails sent by users to pre-existing contacts $f_i$ (see text). The line shows the kernel density estimation of the three yearly datasets pooled together. Most users exchange most of their emails with preexisting contacts. with the maximum at $f_e^{fmax} = 0.90$. **(B)** Distribution of the change of $f_i$, $\Delta f_i(\Delta t) = f_i(t + \Delta t) - f_i(t)$ for $\Delta t = 1, 2$ years (dots and squares, respectively). The lines show the Laplace best fits based on BIC for the two distributions $(P(\Delta f_i) \sim \exp(-|\Delta f_i - \mu|/\sigma)$; $\Delta t = 1$ $\sigma = 0.18$ $\mu = 0.046$; and $\Delta t = 2$ $\sigma = 0.19$ $\mu = 0.062$). Most of the users keep the number of emails sent to preexisting contacts constant in time, and the distributions are quite stable in time despite a slight shift towards larger changes for larger $\Delta t$. **(C)** Comparison between yearly absolute individual change in the fraction of emails sent to preexisting contacts $|\Delta f_{ij}(\Delta t)|_{\text{self}}$ and the typical differences between users $|\Delta f_{ij}|_{\text{ref}} = |f_i(t) - f_j(t)|, \forall j \neq i$. The boxplot shows unambiguously that individual users have a stable communication strategy over time.

doi:10.1371/journal.pone.0146113.g005

To study the stability of each individual's strategy in the long term, we measure the change $\Delta f_i(\Delta t) = f_i(t + \Delta t) - f_i(t)$ at $\Delta t = 1$ and $\Delta t = 2$ years (Fig 5B). First, we find that distributions are stable for fixed $\Delta t$ (Fig B in S4 File). From the distributions, we also observe that most users do not change substantially their communication strategy from year to year. However, 7% of the individuals change their communication strategy by $|\Delta f_i(\Delta t)| > 0.5$, and a small fraction of individuals even change from one end to the other of the communication strategy spectrum.

Despite this variability, we find that, on average, an individual's communication strategy is stable in the long run (Fig 5C). In particular, we compare the absolute individual change $|\Delta f_i(t + \Delta t)|_{\text{self}} = |f_i(t + \Delta t) - f_i(t)|$ with the typical absolute difference between individuals $|\Delta f_{ij}(t)|_{\text{ref}} = |f_i(t) - f_j(t)|, \forall j \neq i$ [3]. We observe that the yearly variation of a user's communication strategy is typically much smaller (even when $\Delta t = 2$ years) than the variation between individuals, confirming the existence of persistent communication strategies even at the scale of several years (see Fig D in S4 File for the analysis of each individual year). By extrapolating the values of $|\Delta f_i(\Delta t)|_{\text{self}}$ as before, we estimate that individual strategies may persist for around seven years.

## Discussion

We have shown that the long-term macro-evolution of email networks follows well-defined distributions, characterized by exponentially decaying log-variations of the weight of social ties and of individuals' social strength. Therefore, the intricate processes of tie formation and decay at the micro-level give rise to macroscopic evolution patterns that are similar to those observed in other complex networks (such as air-transportation or financial networks [18]), as well as in the growth and decay of human organizations [13–17].

The fact that so diverse systems display similar stationary statistical patterns at a macroscopic level (and that these are stable over long periods of time) hints at the existence of universal mechanisms underlying all these processes (such as, for instance, multiplicative processes [16]). Remarkably, together with these statistical regularities, we also observe that individuals have long-lasting social signatures [3] and communication strategies [1, 2], which have a psychological origin, and are unlikely to have a parallel in other systems. Reconciling the universality of the macroscopic evolutionary patterns with the importance of the psychological/microscopic processes should be one of the central aims of future studies about the evolution of social networks.

Last but not least, it will be necessary to understand how the patterns we observe in the evolution of email networks translate into other types of social networks. All existing evidence suggests that email networks (as well as other techno-social networks such as mobile communication networks [31] and online social networks [32]) are good proxies for self-reported friendship-based social networks [26], but more analyses will be necessary to elucidate whether network evolution is also universal. Our finding of stationary and well-defined distributions, and well defined and stable social signatures and communication strategies, suggest that may very well be the case.

## Methods

### Ethics statement

Our work is exempt from IRB review because: i) The research involves the study of existing data-email logs from 2007 to 2010, which the IT service of the organization archived routinely, as mandated by law; ii) The information is recorded by the investigators in such a manner that subjects cannot be identified, directly or through identifiers linked to the subjects. Indeed, subjects were assigned a "hash" by the IT service prior to the start of our research, so that none of the investigators can link the "hash" back to the subject. We have no demographic information of any kind, so de-anonymization is also impossible. Finally, we do not report results for any individual subject (or even for groups of users), but only aggregated results for all users.

### Parameter estimation and model selection for the distribution of logarithmic growth rates

We consider the following functional forms for the distribution of logarithmic growth rates $P(r_\omega^0)$ and $P(r_s^0)$ (see S2 File): i) a Laplace distribution (parameter $\{\sigma_{exp}\}$); ii) a Gaussian distribution (parameter $\{\sigma_G\}$); iii) an asymmetric Laplace distribution (parameters $\{\sigma_{left}, \sigma_{right}\}$); and (iv) the convolution of a Laplace and a Gaussian distribution (parameters $\{\sigma_{exp}, \sigma_G\}$).

We estimate the parameters using maximum likelihood and select the best model using the Bayesian information criterion (BIC) [28] (S2 File). We find that the best model for the distribution $P(r_\omega^0)$ of logarithmic growth rates of connection weight is the convolution of a Laplace and a Gaussian

$$P_{conv}(r_\omega^0 | \sigma_{exp}, \sigma_G) = \int_{-\infty}^{\infty} \frac{e^{-|\rho|/\sigma_{exp}}}{2\sigma_{exp}} \frac{e^{-(r_\omega^0 - \rho)^2/2\sigma_G^2}}{\sigma_G \sqrt{2\pi}} d\rho \ . \tag{5}$$

We find that the best model for the distribution $P(r_s^0)$ of logarithmic growth rates of user strength is Laplace distributed (Eq (3)).

## Supporting Information

**S1 File. Equivalence between the directed and the undirected network of emails.**
(PDF)

**S2 File. Modeling the distribution of logarithmic growth rates.**
(PDF)

**S3 File. Predictability of logarithmic growth rates.**
(PDF)

**S4 File. Social signature and communication strategies.**
(PDF)

## Acknowledgments

We thank the following people for helpful comments and discussions: A. Aguilar-Mogas, F. Massucci, E. Moro, N. Rovira-Asenjo, O. Senan-Campos, T. Vallès-Català, M. Tarrés-Deulofeu.

## Author Contributions

Conceived and designed the experiments: AG-L RG MS-P. Performed the experiments: AG-L RG MS-P. Analyzed the data: AG-L RG MS-P. Contributed reagents/materials/analysis tools: AG-L RG MS-P. Wrote the paper: AG-L RG MS-P.

## References

1. Miritello G, Moro E, Lara R, Martínez-López R, Belchamber J, Roberts SG, et al. Time as a limited resource: Communication strategy in mobile phone networks. Soc Networks. 2013; 35:89–95. doi: 10.1016/j.socnet.2013.01.003

2. Miritello G, Lara R, Cebrian M, Moro E. Limited communication capacity unveils strategies for human interaction. Sci Rep. 2013; 3:1950. doi: 10.1038/srep01950

3. Saramäki J, Leicht E, López E, Roberts SG, Reed-Tsochas F, Dunbar RI. Persistence of social signatures in human communication. Proc Natl Acad Sci USA. 2014; 111:942–947. doi: 10.1073/pnas.1308540110 PMID: 24395777

4. Kossinets G, Watts D. Empirical Analysis of an Evolving Social Network. Science. 2006; 311:88–90. doi: 10.1126/science.1116869 PMID: 16400149

5. McPherson M, Smith-Lovin L, Cook JM. Birds of a Feather: Homophily in Social Networks. Annu Rev Sociol. 2001; 27(1):415–444. doi: 10.1146/annurev.soc.27.1.415

6. Easley D, Kleinberg J. Networks, crowds, and markets: Reasoning about a highly connected world. Cambridge University Press; 2010.

7. Barabási AL. The origin of bursts and heavy tails in human dynamics. Nature. 2005; 435:207–211. doi: 10.1038/nature03459

8. Oliveira JG, Barabási AL. Darwin and Einstein correspondence patterns. Nature. 2005; 437:1251. doi: 10.1038/4371251a PMID: 16251946

9. Malmgren RD, Stouffer DB, Campanharo ASLO, Amaral LAN. On Universality in Human Correspondence Activity. Science. 2009 September; 325(5948):1696–1700. doi: 10.1126/science.1174562

10. Malmgren RD, Ottino JM, Amaral LAN. The role of mentorship in protege performance. Nature. 2010; 465:622–626. doi: 10.1038/nature09040 PMID: 20520715

11. Brockmann D, Hufnagel L, Gaisel T. The scaling laws of human travel. Nature. 2006; 439:462–465. doi: 10.1038/nature04292 PMID: 16437114

12. González MC, Hidalgo CA, Barabási AL. Understanding individual human mobility patterns. Nature. 2008; 453:779–782. doi: 10.1038/nature06958 PMID: 18528393

13. Stanley MHR, Amaral LAN, Buldyrev SV, Havlin S, Leschhorn H, Maass P, et al. Scaling behaviour in the growth of companies. Nature. 1996; 379:804–806. doi: 10.1038/379804a0

14. Amaral LAN, Buldyrev S, Havlin S, Leschorn H, Maass P, Salinger M, et al. Scaling behavior in economics I: empirical results for company growth. J Phys I France. 1997; 7:621. doi: 10.1051/jp1:1997180

15. Amaral LAN, Buldyrev S, Havlin S, Leschorn H, Maass P, Salinger M, et al. Scaling behavior in economics II: modeling of company growth. J Phys I France. 1997; 7:635. doi: 10.1051/jp1:1997181

16. Amaral LAN, Buldyrev SV, Havlin S, Salinger MA, Stanley HE. Power law scaling for a system of interacting units with complex internal structure. Phys Rev Lett. 1998; 80:1385–1388. doi: 10.1103/PhysRevLett.80.1385

17. Plerou V, Amaral LAN, Gopikrishnan P, Meyer M, Stanley HE. Similarities between the growth dynamics of university research and of competitive economic activities. Nature. 1999; 400:433–437. doi: 10.1038/22719

18. Gautreau A, Barrat A, Barthélemy M. Microdynamics in stationary complex networks. Proc Natl Acad Sci USA. 2009; 106:8847–8852. doi: 10.1073/pnas.0811113106

19. Saramäki J, Moro E. From seconds to months: an overview of multi-scale dynamics of mobile telephone calls. Eur. Phys. J. B. 2015; 88:164. doi: 10.1140/epjb/e2015-60106-6

20. Liljeros F, Edling CR, Amaral LA, Stanley HE, Åberg Y. The web of human sexual contacts: Promiscuous individuals are the vulnerable nodes to target in safe-sex campaigns. Nature. 2001; 411:907–908. doi: 10.1038/35082140

21. Liljeros F, Edling CR, Amaral LAN. Sexual networks: implications for the transmission of sexually transmitted infections. Microbes Infect. 2003; 5:189–196. doi: 10.1016/S1286-4579(02)00058-8

22. Balcan D, Colizza V, Gonçalves B, Hu H, Ramasco JJ, Vespignani A. Multiscale mobility networks and the spatial spreading of infectious diseases. Proc Natl Acad Sci U S A. 2009 Dec; 106(51):21484–21489. doi: 10.1073/pnas.0906910106

23. Onnela JP, Saramäki J, Hyvönen J, Szabó G, Lazer D, Kaski K, et al. Structure and tie strengths in mobile communication networks. Proc Natl Acad Sci USA. 2007; 104(18):7332–7336. doi: 10.1073/pnas.0610245104 PMID: 17456605

24. Iribarren JL, Moro E. Impact of human activity patterns on the dynamics of information diffusion. Phys Rev Lett. 2009 Jul; 103(3):038702. doi: 10.1103/PhysRevLett.103.038702 PMID: 19659326

25. Guimerà R, Danon L, Díaz-Guilera A, Giralt F, Arenas A. Self-similar community structure in a network of human interactions. Phys Rev E. 2003; 68:art. no. 065103.

26. Wuchty S, Uzzi B. Human communication dynamics in digital footsteps: A study of the agreement between self-reported ties and email networks. PLOS ONE. 2011; 6(11):e26972. doi: 10.1371/journal.pone.0026972 PMID: 22114665

27. Barrat A, Barthélemy M, Pastor-Satorras R, Vespignani A. The architecture of complex weighted networks. Proc Natl Acad Sci USA. 2004; 101(11):3747–3752. doi: 10.1073/pnas.0400087101

28. Schwarz G. Estimating the dimension of a model. Ann Stat. 1978; 6:461–464. doi: 10.1214/aos/1176344136

29. Breiman L. Random Forests. Mach Learn. 2001; 45(1):5–32. doi: 10.1023/A:1010933404324

30. Dunbar R. The social brain hypothesis. Evol Anthr. 1998; 6(5):178–190. doi: 10.1002/(SICI)1520-6505(1998)6:5%3C178::AID-EVAN5%3E3.0.CO;2-8

31. Eagle N, Pentland A, Lazer D. Inferring friendship network structure by using mobile phone data. Proc Natl Acad Sci USA. 2009; 106(36):15274–15278. doi: 10.1073/pnas.0900282106

32. Dunbar R, Arnaboldi V, Conti M, Passarella A. The structure of online social networks mirrors those in the offline world. Soc. Networks. 2015; 43:39–47. doi: 10.1016/j.socnet.2015.04.005